\documentclass[dvips]{article}
\usepackage{icrctc07}

\title{Discovery of very high-energy $\gamma$-ray emission from the LBL object BL Lacertae}
\shorttitle{Discovery of VHE $\gamma$-ray from BL Lacertae}
\authors{Masaaki~Hayashida$^{1,*}$, Karsten~Berger$^{2}$, Elina~Lindfors$^{3}$, Vincenzo~Vitale$^{4}$, Robert~Wagner$^{1}$ and Eckart~Lorenz$^{1}$ on behalf of the MAGIC collaboration}
\shortauthors{M.~Hayashida et al}
\afiliations{$^1$Max-Planck-Institut f\"ur Physik, D-80805 M\"unchen, Germany\\ $^2$Universit\"at W\"urzburg, D-97074 W\"urzburg, Germany \\ $^3$Tuorla Observatory, Turku University, FI-21500 Piikki\"o, Finland\\ $^4$Universit\`a di Udine, and INFN Trieste, I-33100 Udine, Italy}
\email{mahaya@mppmu.mpg.de}

\abstract{The low-frequency peaked BL Lac (LBL) object BL Lacertae was observed with the MAGIC telescope from 2005 August to December  (22.2 hr), and from 2006 July to September (26.0 hr). A very high energy (VHE) $\gamma$-ray signal was discovered with a 5.1 $\sigma$ excess in the 2005 data. Above 200 GeV, an integral flux of $(0.6\pm0.2)\times10^{-11}~{\rm cm}^{-2}~{\rm s}^{-1}$ was measured, corresponding to approximately 3\% of the Crab flux. The differential spectrum between 150 and 900 GeV is rather steep, with a photon index of $-3.6\pm0.5$. 
The light curve shows no significant variability during the observations in 2005.
For the first time a clear detection of VHE $\gamma$-ray emission from an LBL object was obtained.
The 2006 data show no significant excess. This drop in flux follows the observed trend in optical activity.}

\begin{document}
\maketitle

\section{Introduction}

  BL Lacertae (1ES2200+420, z=0.069, \cite{Miller78}) is the historical prototype of a class of powerful $\gamma$-ray emitters: "BL Lac objects". It belongs to an active galactic nucleus (AGN) subclass in which the jet is aligned very close to our line of sight. 
Electromagnetic emission from this class of sources can be observed from the radio up to very high energy (VHE) $\gamma$-rays $(E>$100 GeV), with spectral energy distributions (SEDs) characterized by a two-bump structure. 
The lower energy bump is produced by synchrotron radiation of relativistic electrons, while the higher energy bump originates from inverse Compton (IC) scattering of the same 
synchrotron photons (synchrotron self-Compton scattering, SSC) or external photons either from the broad-line emission region or from the accretion disc (external inverse Compton scattering, EC)~\cite{Ghi98}. 
When the synchrotron emission peak is located in the sub-millimeter to optical band, the objects are classified as ''Low-frequency peaked BL Lacs" (LBLs), whereas in ''High-frequency peaked BL Lacs'' (HBLs) the synchrotron peak is located at UV to X-ray energies~\cite{Pad95}. 
It should be noted that other models, based e.g.\,on the acceleration of hadrons~\cite{Man93, Muc01} could also explain the SEDs of BL Lac objects.

At this conference, in total 19 extragalactic VHE $\gamma$-ray sources were reported~\cite{Jim}. Except for three sources (BL Lacertae~[this work], 3C279~\cite{Teshi} and M87~\cite{M87}) all other sources are classified as HBLs.
BL Lacertae is classified as an LBL object with a synchrotron peak frequency of $2.2 \times 10^{14}$ Hz~\cite{Sam99}.
Before this work, no VHE $\gamma$-ray emission from any LBLs had been confirmed.

\section{Observations and Results}

The MAGIC telescope, located at the Canary Island La Palma (N28.2$^{\circ}$, W17.8$^{\circ}$, 2225 m a.s.l.), is an Imaging Atmospheric Cherenkov Telescope (IACT) with a 17m diameter dish.
The telescope parameters and performance are described in \cite{Crab}.

Optical $R$-band observations were provided by the Tuorla Observatory Blazar Monitoring Program
with the 1.03 m telescope at the Tuorla Observatory, Finland, and the 35 cm KVA telescope on La Palma, Canary Islands. 

Radio observations were also performed with UMRAO
at 4.8, 8.0 and 14.5 GHz, and at 37 GHz with the Mets\"ahovi Radio Observatory.

   \begin{figure}
   \centering
   \includegraphics[width=6.6cm, clip]{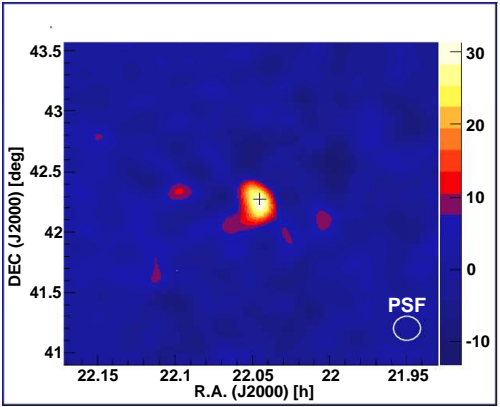}
                      \caption{Sky map of the region around the position of BL Lacertae (black cross) for reconstructed $\gamma$ events with $>$ 350 photo-electrons (corresponding to an energy threshold of about 200 GeV) from the 2005 observations.}	
         \label{sky}
   \end{figure}
%
%

\subsection{The 2005 data}
BL Lacertae (R.A.$22^{\rm h}02^{\rm m}43.3^{\rm s}$, decl.$+42^{\circ}16'40''$  ; J2000.0) was observed for 22.2 hr from 2005 August until December 2005 by MAGIC. The telescope at that time was pointing directly onto the object, recording so-called ON-data. The background was estimated from observations of regions where no $\gamma$-rays are expected, so-called OFF-data, which were taken with sky conditions similar to those of ON-data.
After the data quality selection, 
the remaining ON-data corresponded to 17.8 hr, while the OFF-data corresponded to 57.2 hr. Image parameters~\cite{Hillas} were calculated and compared for the ON and OFF data in order to check their consistency; excellent agreement was found. 
The details of the analysis procedure can be found in~\cite{Crab, MAGIC07b, MAGIC07c}.

   \begin{figure}
   \centering
\includegraphics[width=7cm, clip]{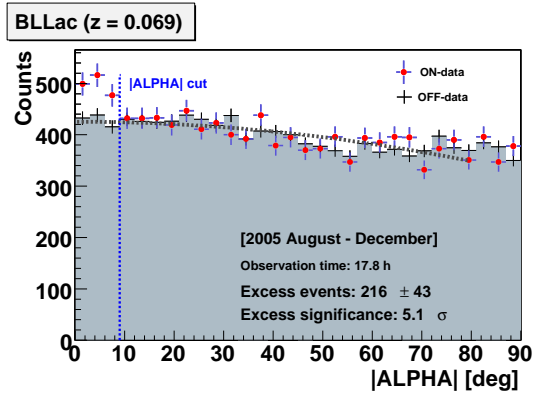}
   \caption{ALPHA-distribution of the 2005 data. 
   The vertical line indicates the ALPHA selection condition ($< 9^{\circ}$), which yielded a total excess of 216 events at a significance level of 5.1 $\sigma$.
   }
   \label{alpha}
   \end{figure}

\begin{figure*}
\centering
\includegraphics [width=13.0cm, clip]{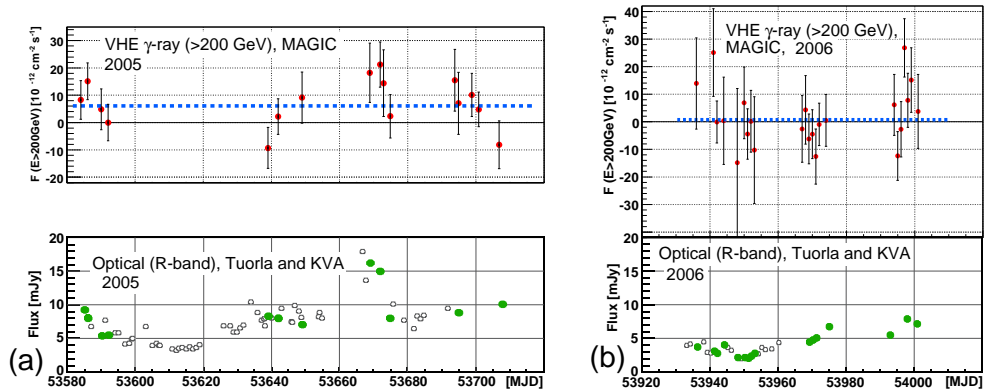}
 \caption{[\textbf{Upper}] Light curve of VHE $\gamma$-ray ($>$ 200 GeV) flux as measured with the MAGIC telescope. A dotted horizontal line represents the average flux. [\textbf{Lower}] Light curve of optical ($R$-band) flux as measured with the Tuorla and the KVA telescope. The filled points represent the optical flux when simultaneous MAGIC observations were carried out. [\textbf{left}] (a) The 2005 observation. [\textbf{right}] (b) The 2006 observation.
              }
         \label{LC}
\end{figure*}


 Figure~\ref{sky} shows the local sky map for the 2005 $\gamma$-ray candidates. Only events $>$ 350 photo-electrons were used, corresponding to an energy threshold of about 200 GeV.

In Figure~\ref{alpha}, the ALPHA distribution is shown. 
An excess of 216 events over 1275.6 normalized background events yields a significance of 5.1 $\sigma$ for data above 350 photo-electrons. 

The 2005 VHE $\gamma$-ray and the optical light curves are shown in Figure~\ref{LC} (a). 
No significant evidence of flux variability in VHE $\gamma$-rays was found in the 2005 data while the optical light curve shows a flare around the end of October 2005 (MJD $\sim$53670). 
The derived average integral flux is $F(E>200\ {\rm GeV})= (0.6 \pm 0.2) \times 10^{-11}\ {\rm cm}^{-2}\ {\rm s}^{-1}$ $ (\chi^{2}/\rm{dof} = 16.3/15)$, which corresponds to about 3\% of the Crab Nebula flux as measured by the MAGIC telescope~\cite{Crab}.

\begin{figure}
   \centering
   \includegraphics[width=7cm]{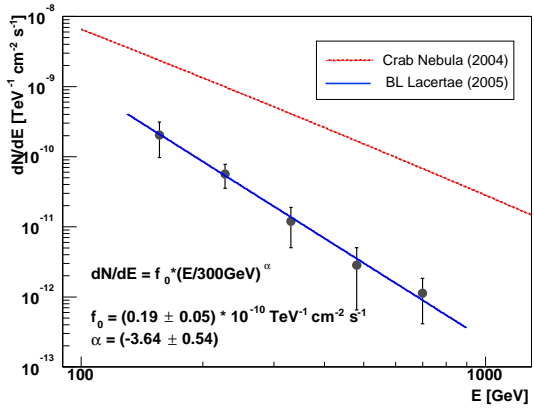}
                       \caption{Differential energy spectrum of the 2005 BL Lacertae data. The solid line represents a power-law fit to the measured spectrum. The fit parameters are listed in the figure. For comparison, the line fitted to the measured MAGIC CRAB spectrum using a power law with a changing photon index is shown by a dashed line~\cite{Wag05}.}
         \label{sp}
   \end{figure}

The reconstructed differential energy spectrum (Fig.~\ref{sp}) is well described by a simple power law:
\begin{equation}\label{equ1}
 \frac{dN}{dE} = (1.9 \pm 0.5) \times 10^{-11} \left( \frac{E}{300\ {\rm GeV}}\right)^{-3.6\pm0.5} 
\end{equation}
[${\rm TeV}^{-1}\ {\rm s}^{-1}\ {\rm cm}^{-2}$]. Because of the relatively steep slope, the systematic errors are estimated to be $\sim$ 50\% for the absolute flux level and 0.2 for the spectral index. 

\subsection{The 2006 data}
Follow-up observations were carried out from 2006 July to September for 26.0 hr, using the so-called wobble mode~\cite{Dau97}, where the object was observed at an $0.4^{\circ}$ offset from the camera center; 25.0 hr of the data passed all quality selection for the analysis. 
Figure~\ref{LC} (b) shows the VHE $\gamma$-ray and the optical light curves in 2006. 
No significant VHE $\gamma$-ray excess could be found in the 2006 data.

\section{Discussion}

The Whipple 10m telescope observed BL Lacertae for 39.1 hr in 1995 and derived a flux upper limit above 350 GeV at 3.8\% of Crab~\cite{Hor04}. HEGRA derived an upper limit above 1.1 TeV at 28\% of Crab with 26.7 hr observation~\cite{Aha04}. These upper limits are consistent with our results. On the other hand, Neshpor et al.~\cite{Nes01} claim VHE $\gamma$-ray detection using the GT 48 telescopes of the Crimean Astrophysical Observatory from data taken in summer 1998. The reported integral $\gamma$-ray flux is $F(E>1\ {\rm TeV}) = (2.1 \pm 0.4) \times 10^{-11}\ {\rm cm}^{-2}\ {\rm s}^{-1}$, which is two orders of magnitude higher than the extrapolated value of this study. During the same period, 1998 July to August, no significant signal was found by HEGRA, and their reported flux upper limit in the same energy band is 7 times lower then the Crimean result~\cite{Kra03}. The VHE $\gamma$-ray detection reported by \cite{Nes01} cannot be explained without a remarkably huge and a very rapid flux variation offset by a few hours in consecutive nights from the HEGRA observation. In the case of a leptonic origin of the $\gamma$-ray emission, such a flare would possibly coincide with high activity in the optical as in the outburst of 1997 July, when the increase in flux was observed both in the optical and X-ray to $\gamma$-ray bands~\cite{Blo97, Tan00}. However, no increased optical activity was detected during the Crimean observation period ($m_p = 13.5-14.6$, while $13.0-14.6$ in 2005). 

 In Figure~\ref{LC}, filled circles in the optical light curve represent simultaneous observations with the MAGIC telescope, accepting a $\pm$1 day offset with respect to the MAGIC observations. 
In 2005, there are 12 out of 16 nights with simultaneous observations (average flux: 9.2 mJy), while 16 out of 23 nights in 2006 (average flux: 4.2 mJy) have coinciding observations. 
The absence of a significant excess of VHE $\gamma$-rays in the 2006 data indicates that the VHE $\gamma$-ray flux in 2006 was significantly lower than the flux level in 2005. 
In summary our results show similar tendencies both in the optical and the VHE $\gamma$-ray flux variations. Similarly, the $\gamma$-ray activity seen by EGRET observations in 1997 showed a strong correlation with optical activity. Such a correlation is expected from leptonic origin scenarios~\cite{Blo97}.

%
%

\begin{figure}
   \centering
   \includegraphics[width=7.1cm]{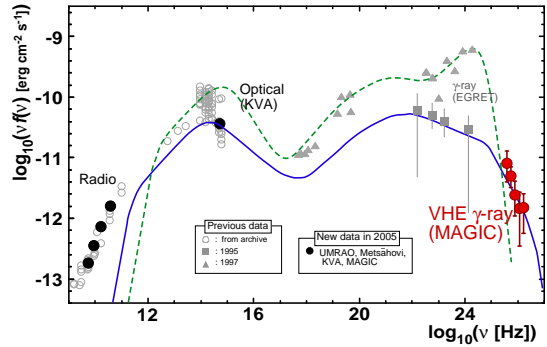}
           \caption{SED for BL Lacertae. Black filled circles represent simultaneous 2005 data of MAGIC, KVA and radio data taken by UMRAO and Mets\"ahovi. 
Gray color points describe old measurements (see detail in the inlay). 
The lines are taken from \cite{Rav02}. The solid line represents one-zone SSC model for 1995 data, the dotted line is produced with SSC and EC components for 1997 flare data.}
         \label{SED}
   \end{figure}

 Figure~\ref{SED} shows the SED of BL Lacertae with results of this work and some previous data and model calculations by \cite{Rav02}. The VHE $\gamma$-ray points are corrected for the extragalactic background light (EBL) absorption using the "Low" EBL model of \cite{Kne04}. Our optical and VHE $\gamma$-ray points agree well with the solid line, which was derived using a one-zone SSC model on the 1995 data, whereas some deviations can be seen from the dotted line, which describes the 1997 flare data and involves SSC as well as EC components~\cite{Rav02}. 
To describe our result such an additional EC component is not necessarily required.

BL Lacertae is the first LBL with a clear detection of VHE $\gamma$-ray emission. The results of this work indicate that VHE $\gamma$-ray observations during times of higher optical states can be more efficient.
Due to the observed steep spectrum, lowering the energy threshold of IACTs (e.g. the upcoming MAGIC-II project), would significantly increase the detection prospects for this new class.

%

\section{Acknowledgements}
We would like to thank the IAC for the excellent working conditions at the ORM. The support of the German BMBF and MPG, the Italian INFN, the Spanish CICYT, the Swiss ETH and the Polish MNil is gratefully acknowledged.
We thank A. L\"ahteenm\"aki (Mets\"ahovi), H. Aller and M. Aller (UMRAO) for providing the radio data.



\bibliography{icrc0946}

\begin{thebibliography}{10}

\bibitem{Aha04}
F.~Aharonian et~al.
\newblock {\em A\&A}, 421:529, 2004.

\bibitem{M87}
F.~Aharonian et~al.
\newblock {\em Science}, 314:1424, 2006.

\bibitem{Crab}
J.~Albert et~al.
\newblock {\em ApJ}, 2007.
\newblock submitted (astro-ph/0705.3244).

\bibitem{MAGIC07b}
J.~Albert et~al.
\newblock {\em ApJ}, 654:L119, 2007.

\bibitem{MAGIC07c}
J.~Albert et~al.
\newblock {\em ApJL}, 666, 2007.
\newblock in press (astro-ph/0703084).

\bibitem{Blo97}
S.~D. Bloom et~al.
\newblock {\em ApJ}, 490:L145, 1997.

\bibitem{Dau97}
A.~Daum et~al.
\newblock {\em Astropart. Phys.}, 8:1, 1997.

\bibitem{Ghi98}
G.~Ghisellini et~al.
\newblock {\em MNRAS}, 301:451, 1998.

\bibitem{Hillas}
A.~M. Hillas.
\newblock {\em Proc. 29th Int. Cosmic Ray Conf. (La Jolla)}, 3:445, 1985.

\bibitem{Jim}
J.~Hinton.
\newblock 2007.
\newblock Rapporteur talk (OG 2.1 - 2.4) in this conference.

\bibitem{Hor04}
D.~Horan et~al.
\newblock {\em ApJ}, 603:51, 2004.

\bibitem{Kne04}
T.~M. Kneiske et~al.
\newblock {\em A\&A}, 413:807, 2004.

\bibitem{Kra03}
D.~Kranich.
\newblock {\em ASP Conf. Proc.}, 299:3, 2003.

\bibitem{Man93}
K.~Mannheim.
\newblock {\em A\&A}, 269:76, 1993.

\bibitem{Miller78}
J.~S. Miller et~al.
\newblock {\em ApJ}, 219:L85, 1978.

\bibitem{Muc01}
A.~M{\"u}cke and R.~J. Protheroe.
\newblock {\em AsrtoPart. Phys.}, 15:121, 2001.

\bibitem{Nes01}
Y.~I. Neshpor et~al.
\newblock {\em Astron.~Rep.}, 45:249, 2001.

\bibitem{Pad95}
P.~Padovani and P.~Giommi.
\newblock {\em ApJ}, 444:567, 1995.

\bibitem{Rav02}
M.~Ravasio et~al.
\newblock {\em A\&A}, 383:763, 2002.

\bibitem{Sam99}
R.~M. Sambruna et~al.
\newblock {\em ApJ}, 515:140, 1999.

\bibitem{Tan00}
C.~Tanihata et~al.
\newblock {\em ApJ}, 543:124, 2000.

\bibitem{Teshi}
M.~Teshima et~al.
\newblock 2007.
\newblock these proceedings (\#1019).

\bibitem{Wag05}
R.~M. Wagner et~al.
\newblock {\em Proc. 29th Int. Cosmic Ray Conf. (Pune)}, 4:163, 2005.

\end{thebibliography}
\bibliographystyle{plain}

\end{document}